\documentclass[aps,prb,amsmath,amssymb
,showpacs
,twocolumn
,floatfix
]{revtex4}
\usepackage{graphicx}
\usepackage{dcolumn}
\usepackage{bm}
\begin{document}
\title{\bf Block-Diagonalization and f-electron Effects in Tight-Binding Theory}

\author{Matthew D. Jones}
\affiliation{Department of Physics and Center for Computational Research,\\
University at Buffalo, The State University of New York,
Buffalo, NY 14260}
\email{jonesm@ccr.buffalo.edu}

\author{R. C. Albers}
\affiliation{Theoretical Division, Los Alamos National Laboratory, \\
Los Alamos, NM 87545}
\email{rca@lanl.gov}

\date{\today}
\begin{abstract}

We extend a tight-binding total energy method to include
$f$-electrons, and apply it to the study of the structural and elastic
properties of a range of elements from Be to U.  We find that the
tight-binding parameters are as accurate and transferable for
$f$-electron systems as they are for $d$-electron systems.  In both
cases we have found it essential to take great care in constraining
the fitting procedure by using a block-diagonalization procedure,
which we describe in detail.

\end{abstract}

\maketitle


\section{Introduction}

In order to use atomistic modeling to predict materials properties it
is necessary to accurately determine the forces between the atoms in a
solid.  Most of the established molecular dynamics simulations involve
either pair potentials, embedded-atom potentials, or, for covalent
materials like Si, some classical potentials with additional
bond-angle information.  Unfortunately, these methods have often
proved inadequate for complex transition-metal and $f$-electron
systems, which have important bond-bending forces that seem difficult
to capture in any simple way.  A promising approach for such systems
is tight-binding (TB), since it automatically builds in the
quantum-mechanical bonding that conventional first-principles
local-density approximation (LDA) or gradient corrected (GGA)
band-structure calculations can very accurately determine.

In this work we report on our progress to extend the recent TB total
energy model that was developed at the U.S. Naval Research
Laboratory(NRL); this has been successfully applied to semiconductors
and both simple and transition
metals\cite{cohen94,mehl96,yang98,papacon86,mehl98}.  We have extended
and tested the NRL-TB scheme to include $f$-electrons, which makes
possible the exploration of materials with important $f$
characteristics, such as the light actinides.  We also describe our
own specific methodology for our fitting procedures, which is based on
the NRL procedures, that we believe should produce parameterizations
that are highly accurate and should be well suited for molecular
dynamics simulations, although we report no such simulations in this
paper.  In the rest of the paper, we first briefly recapitulate our
modifications to the NRL approach, and describe our experience in
obtaining high quality TB parameters. We test this approach for an s-p
bonded material (Be), $d$-electron materials with nearly filled and
partially filled $d$-bands (Cu and Nb), and finally for a partially
filled $f$-band system (U), an element of significant technological
importance, which has complex structural properties.

One of the major obstacles in atomistic methods for materials modeling
is the question of transferability, which arises from the
semi-empirical nature of the atomistic forces, which involve fitting
parameters to either experimental data or to theoretical calculations.
The problem of transferability is the question of how well such
fitting parameters will work for geometries of atoms that are
different from those to which they were fit.  This question is
especially important for molecular dynamics simulations, where the
atoms are free to move wherever the atomic forces push them.  It is
also needed for determining ground-state structures --- the structure
that will minimize the total energy of the system --- or for
calculating various defect structure (vacancies, interstitials, and
more complex defects) that are expensive for first-principles methods
(LDA calculations).  In this paper we will show that the TB method
works for a wide variety of different atomic environments (e.g., for
crystal structures with very different numbers of near neighbors) and
over a large atomic volume range.

Another qualitative feature of atomistic potentials that is required
for accurate molecular dynamics and other atomistic simulations is to
parameterize self-consistency or local environmental effects.  To
understand what we mean by this statement, consider first-principles
electronic-structure calculations for the total energy of a solid as a
function of the atomic positions.  Such calculations depend on
determining a self-consistent potential for the electrons for every
atomic structure considered.  An electronic potential for any given
atom depends importantly on the local arrangement of nearby atoms.
This implies that any force parameterization that successfully mimics
the electronic-structure calculation also senses changes in the local
atomic environments.  This concept has been a major driving force for
adding local environmentally sensitive terms in atomic force models.
The NRL-TB parameterization also contains such effects for onsite
terms (see below).  One strong test of local atomic rearrangements is
the number of near-neighbors, which should strongly perturb the local
charge density around any given atom and hence exacerbate
self-consistency effects.  For this reason we have focused our fitting
procedures for TB parameters on crystal structures with widely varying
numbers of nearest neighbors.  As we will describe below, the NRL
parameterization appears to have adequate flexibility to successfully
pass this fairly stringent test.

All empirical force models use some procedure to fit their parameters
(usually some variation on a least-squares fit to either experimental
or theoretical data).  As more parameters are added, perhaps to better
describe bond-bending forces or approximate environmental effects,
overall properties such as the total energy often become less
sensitive to the parameters.  What is actually worse is that different
parameters can sometimes compensate for each other so that parameter
space becomes a vast multidimensional landscape where the minimizing
function often has large numbers of weak minima and it is difficult to
find the physically relevant true global minimum that provides maximum
transferability.  Since there is often a choice of different
parameterizations, it can be difficult to know whether the lack of
success in fitting properties is due to a parameterization that has
inadequate flexibility or a failure to find and lock onto the best
choice of parameters when wandering through the enormous
parameterization space during the minimization process.

In our TB fits, we have found it to be enormously important to add
constraints to the fitting procedure in order to force the parameters
to be near the physically relevant portion of parameter space.  Once
this is done, any minimization procedure seems to improve the
minimizing function as well as the transferability.  Using physically
motivated constraints appears to remove the huge degeneracy of
multiply compensating parameters and to help focus the
parameterization onto the desired solution.  In the case of TB, as we
will discuss below, the key constraint is symmetry.  If used properly,
it appears to force parameter space into highly transferable
parameterizations.

\section{TB Method}

Theoretical justification for the tight binding method
rests upon the division of the total energy of the system
into a repulsive pairwise term and contributions from the
valence band structure,
\begin{equation}
E_T = E_{rep} + E_b,
\label{eq:decomp}
\end{equation}
itself an approximation to the Harris-Foulkes\cite{harris85,foulkes89}
functional, which can be derived from Kohn-Sham density functional 
theory\cite{hohenberg64,kohn65} (DFT),
\begin{widetext}
\begin{equation}
{\cal E} [n] = \sum_{i,v} \epsilon_{i,v} - E_H[n_v]-\int n_v(\epsilon_{xc}[n_v]
               -V_{xc}[n_v])+\frac{1}{2}\sum_a \sum_{b\ne a} Z_{va} Z_{vb}/
                \left| {\bf R}_a-{\bf R}_b \right| .
\label{eq:hff}
\end{equation}
\end{widetext}
In Eq. \ref{eq:hff}, the subscript $v$ denotes the fact that we
are dealing only with the valence electrons whose density is given by
$n_v$, and the first term on the right hand side is the band term - a sum
over eigenvalues arising from a Schr\"odinger-like equation. $E_H$ denotes
the Hartree energy, and $\epsilon_{xc}$ and $V_{xc}$ the exchange and
correlation energy and potential, respectively. A simple decomposition of
this functional into pairwise and bonding terms (Eq. \ref{eq:decomp}) can
be justified\cite{harris85,foulkes89} when terms involving more than two 
atomic centers
are ignored (see Eq. \ref{eq:hamsum} below), and the input charge density
is atomic-like. The Slater-Koster\cite{slater54} method then consists of
solving the secular equation,
\begin{equation}
H\psi_{i,v}=\epsilon_{i,v} S\psi_{i,v},
\label{eq:sec}
\end{equation}
for the single particle eigenvalues and orbitals, under the restrictions:
terms involving more than two centers are ignored, terms where the
orbitals are on the same atomic site are taken as constants, and
the resulting reduced set of matrix elements are treated as variable
parameters.  Note that Eq. \ref{eq:sec} includes the overlap matrix,
$S$, to take into account that the basis functions need not be orthogonal.
To better illustrate these restrictions, we write the
Hamiltonian including the labels for orbitals having generic quantum
numbers $\alpha,
\beta$ localized on atoms $i,j$, where the effective potential is assumed
to be spherical, and can be represented as a sum over atomic centers,
\begin{equation}
H_{\alpha i,\beta j} = \left\langle \alpha,i \left| -\nabla^2
                        +\sum_k V^{\rm eff}_k \right| \beta,j \right\rangle,
\label{eq:hamsum}
\end{equation}
which we further decompose into ``on-site'' and ``inter-site'' terms,
\begin{equation}
H_{\alpha i,\beta j} = e_\alpha \delta_{\alpha\beta} \delta_{ij}
        + E_{\alpha i,\beta j}(1-\delta_{ij}),
\end{equation}
where the on-site terms, $e_\alpha$, represent terms in which two
orbitals share the same atomic site, and, for $j \ne i$,
\begin{align}
E_{\alpha i,\beta j} &= \sum_n e^{i \textbf{k}\cdot\left(\textbf{R}_n+
\textbf{b}_j-\textbf{b}_i\right)} \nonumber \\
 & \times \int d\textbf{r} \psi_\alpha\left(
        \textbf{r}-\textbf{R}_n-\textbf{b}_i\right) H
        \psi_\beta\left( \textbf{r}-\textbf{b}_j\right), \label{eq:Eint}\\
S_{\alpha i,\beta j} &= \sum_n e^{i \textbf{k}\cdot\left(\textbf{R}_n+
\textbf{b}_j-\textbf{b}_i\right)} \nonumber \\
 & \times \int d\textbf{r} \psi_\alpha\left(
        \textbf{r}-\textbf{R}_n-\textbf{b}_i\right)
        \psi_\beta\left( \textbf{r}-\textbf{b}_j\right), \label{eq:Sint}
\end{align}
are the remaining energy integrals involving orbitals located on different
atomic sites. In Eqs. \ref{eq:Eint}-\ref{eq:Sint} we have used translational
invariance to reduce the number of sums over bravais lattice points
$\{\textbf{R}_n\}$, and the $\textbf{b}_i$ denote atomic basis vectors
within the repeated lattice cells.
Orthogonal TB (OTB) treatments do not use Eq. \ref{eq:Sint}, and thus have
fewer parameters, but we have found a non-orthogonal (NOTB) approach to
be more accurate, consistent with previous TB work\cite{cohen97}.
All of the results that we report in this paper use a NOTB model.
Note that terms which have both
orbitals located on the same site, but the potential on other sites have
been ignored - these contributions are typically taken to be
``environmental'' corrections to the on-site terms, and are not accounted
for in the usual Slater-Koster formalism (although such corrections have
been explored, for example, see Ref. 11 and references therein). The two 
center approximation then consists of ignoring additional terms in the 
inter-site contributions in which the effective potential does not lie on 
one of the atomic sites. This approximation\cite{slater54} is not necessarily 
very accurate (see, for example, the comparison between two and three-center
fits in Ref.~\onlinecite{papacon86}), but it is quite often used due to the 
enormous
simplification of the overall TB method; perhaps one can view TB as
having the right functional form and the choice of parameters allow some
correction for any errors due to neglected terms.  Once this approximation
has been made, well justified or not, the inter-atomic ($i\ne j$) matrix
elements reduce to a simple sum over angular functions and functions
which depend only upon the magnitude of the distances between atoms,
\begin{eqnarray}
H_{\alpha i,\beta j} = \sum h_{ll'm}(r_{ij}) G_{ll'm}(\Omega_{i,j}),\label{eq:mat2}\\
S_{\alpha i,\beta j} = \sum s_{ll'm}(r_{ij}) G_{ll'm}(\Omega_{i,j}),
\end{eqnarray}
where we have now adopted the usual convention of using
the familiar $l,m$ angular momentum quantum numbers,
and the axis connecting the atoms is the quantization axis.
The basis set used for the $\alpha$ and $\beta$ quantum states are
the cubic harmonics\cite{vonderlage47} 
whose functional forms are given by (with appropriate normalization factors)
\begin{equation}
\label{eq:TBbasis}
\begin{array}{ll}
| s\pm \rangle &= \sqrt{1/4\pi} |\pm\rangle \nonumber \\
| p_1\pm \rangle &= \sqrt{3/4\pi} f_p(r) x |\pm\rangle \nonumber\\
| p_2\pm \rangle &= \sqrt{3/4\pi} f_p(r) y |\pm\rangle \nonumber\\
| p_3\pm \rangle &= \sqrt{3/4\pi} f_p(r) z |\pm\rangle \nonumber\\
| d_1\pm \rangle &= \sqrt{5/16\pi} f_d(r) xy |\pm\rangle \nonumber\\
| d_2\pm \rangle &= 2\sqrt{15/16\pi} f_d(r) yz |\pm\rangle \nonumber\\
| d_3\pm \rangle &= 2\sqrt{15/16\pi} f_d(r) zx |\pm\rangle \nonumber\\
| d_4\pm \rangle &= \sqrt{15/16\pi} f_d(r) (x^2-y^2) |\pm\rangle \nonumber\\
| d_5\pm \rangle &= \sqrt{5/16\pi} f_d(r) (3z^2-r^2) |\pm\rangle \nonumber\\
| f_1\pm \rangle &= 2\sqrt{105/16\pi} f_f(r) xyz |\pm\rangle \nonumber\\
| f_2\pm \rangle &= \sqrt{7/16\pi} f_f(r) x(5x^2-3r^2) |\pm\rangle \nonumber\\
| f_3\pm \rangle &= \sqrt{7/16\pi} f_f(r) y(5y^2-3r^2) |\pm\rangle \nonumber\\
| f_4\pm \rangle &= \sqrt{7/16\pi} f_f(r) z(5z^2-3r^2) |\pm\rangle \nonumber\\
| f_5\pm \rangle &= \sqrt{105/16\pi} f_f(r) x(y^2-z^2) |\pm\rangle \nonumber\\
| f_5\pm \rangle &= \sqrt{105/16\pi} f_f(r) y(z^2-x^2) |\pm\rangle \nonumber\\
| f_5\pm \rangle &= \sqrt{105/16\pi} f_f(r) z(x^2-y^2) |\pm\rangle,
\end{array}
\end{equation}
where $f_l(r)=1/r^l$, and $|\pm\rangle$ denotes the spin-state.
The selection of these particular functions is not accidental,
as they are chosen to specifically possess the various irreducible
representations of the cubic point group ${\cal O}_h$.
In principle we are free to choose any set of orthogonal functions
as our basis, but it is beneficial to choose the set that best
reflects the symmetry properties of the system under study.
Other choices exist\cite{sharma79}, but they do not provide nearly
as transparent a framework for modeling, fitting and understanding the
various parameters.

The Slater-Koster tables for the $sp^3d^5$ matrix elements can be
found in standard references\cite{harrison80}, and we have used
the tabulated results of Takegahara {\it et al.}\cite{takegahara80} for the
additional matrix elements involving $f$-electrons.
Typical TB applications are then reduced to using TB as an interpolation
scheme; the matrix elements ($h_{ll'm}$, $s_{ll'm}$ (if used) and 
$e_\alpha$) are
determined by fitting to {\it ab-initio} calculated quantities such as the
total energy and band energies.
The NRL-TB total energy method was an innovation in that the energy bands
used in the TB fits were shifted such that the integrated band energy
was the total energy,
\begin{equation}
\epsilon_b'=(\epsilon_b-E_b+E_{tot})/N_v,
\end{equation}
where $\epsilon_b$ are the unshifted energy band values, 
$E_b=\sum_b \epsilon_b$ the total band energy, and $E_{tot}$
the computed total energy.
The NRL-TB method imposes a simple functional form on the inter-site
matrix elements,
\begin{eqnarray}
h_{ll'm}(r) &= \left( a_{ll'm}+b_{ll'm} r\right) \exp\left(-c_{ll'm}^2 r\right)
  f_c(r), \\
s_{ll'm}(r) &= \left( \bar{a}_{ll'm}+\bar{b}_{ll'm} r\right) \exp\left(-\bar{c}_{ll'm}^2 r\right)
  f_c(r), 
\label{eq:distdep}
\end{eqnarray}
where $f_c=1/(1+\exp(2*(r-r_0)))$ is a multiplicative factor included to
ensure a smooth cutoff with increasing distance.  Our applications have used
$r_0=13.5$ bohr radii for Cu and Nb, $r_0=10$ for Be, and $r_0=12.0$ for U.
The on-site terms include the environment dependence necessary to
allow the parameterization to be applied to other structures not included 
in the TB database,
\begin{equation}
e_\alpha = e_\alpha^0+e_\alpha^1\rho^{2/3}+e_\alpha^2\rho^{4/3}
+e_\alpha^3\rho^{2}.
\label{eq:onsite}
\end{equation}
The $\rho$ function is intended to be a crude measure of the atomic
density (providing a correlation to the number of atomic neighbors),
\begin{equation}
\rho = \sum_{i\ne j} \exp\left(-\lambda^2 r_{ij}\right) f_c(r_{ij}),
\end{equation}
where $r_{ij}$ is the interatomic distance.
Note that, for a NOTB model in our $sp^3d^5f^7$ basis, this TB model
has $1+16+2*(3*20)=137$ free parameters, whose determination is the
key to an accurate representation.

\subsection{Fitting the Parameters}

\begin{figure}[htb]
\includegraphics[scale=0.5]{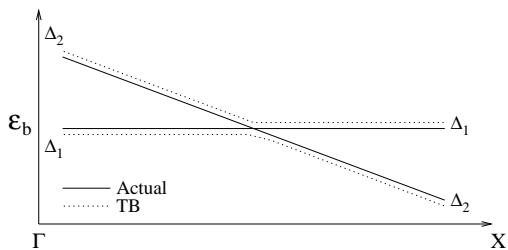}
\caption{\label{fig:bd}
A schematic representation of the band ordering problem when trying to
obtain high quality TB fits.  Here the actual bands correspond to two
different irreducible representations, labeled $\Delta_1$ and
$\Delta_2$.  The TB fits, when the symmetry is ignored and only the
energy ordering is used, can miss essential features, like the band
crossing pictured here.  In this figure the $\Delta$ direction in the
Brillouin zone is the abscissa, while the ordinate is increasing
energy.}
\end{figure}

In order to determine such a large set of parameters, the usual NRL-TB
approach\cite{cohen94,mehl96,yang98,papacon86,mehl98} involves
performing a non-linear least squares minimization, fitting to the
energy bands and total energies calculated from density functional
theory.  The energy bands are fitted over a set of points in the
irreducible wedge of the first Brillouin zone (IBZ).  In practice we
have found it absolutely necessary to take maximum advantage of the
symmetry present in order to impose constraints on the fitting
process.  We exploit the symmetry of all possible high-symmetry points
and lines in the first Brillouin zone in order to reduce the
possibility that the fitting process can easily mistake the ordering
of the energy bands.  We have decomposed the TB wavefunction at each
high symmetry point and line in terms of the symmetry-adapted TB basis
functions\cite{cornwell69}, which allows us to block-diagonalize the
Hamiltonian and overlap matrices, and determine the eigenvalues
corresponding to each irreducible representation, which avoids
possible confusion as to the band ordering.  This process is
illustrated schematically by Figure \ref{fig:bd}.
Figure \ref{fig:bd} shows the perils of fitting using a simple energy
ordering scheme for the bands - unless the irreducible representation is
identified and constrained for each band, the actual fitting errors
are misidentified.

Let us describe the scheme mathematically and follow up with a simple
example.
Along symmetry lines and planes in the Brillouin zone, there exist symmetry
operations for each $k$ that form a group.  For each such group the irreducible
representations $\Gamma^p$ can be found, and can be used to block-diagonalize
the secular equation,
\begin{equation}
D(p)_{ji} = \left\langle \phi^p_j \left| H \right| \phi^p_i \right\rangle
 - E \left\langle \phi^p_j | \phi^p_i \right\rangle,
\label{eq:blockD}
\end{equation}
where $\textbf{D}(p)$ now takes on the ``block-form'', and the symmetry
reduced eigenvalues can be found by solving
\begin{equation}
\left|
\begin{array}{cccc}
   D(1) & 0 & 0 & \ldots \\
   0 & D(2) & 0 & \ldots \\
   0 & 0 & D(3) & \ldots \\
  \vdots & \vdots & \vdots & \vdots\\
\end{array}
\right| = 0,
\label{eq:blockD2}
\end{equation}
or simply $\left| \textbf{D}(p) \right| = 0$ for every $p$.

The resulting set of symmetry adapted basis functions are specific linear
combinations of the original cubic harmonic basis functions 
(Eq. \ref{eq:TBbasis}), which can be found, in practical terms, by applying
projection operators of the irreducible representations to the original
basis functions.  In our fits, we require that only the correct symmetry
adapted tight-binding basis functions can be used for fitting each energy
eigenvalue for high symmetry $k$ points and directions in the Brillouin zones
for the various cubic crystal structures.
In this way we fit not only the energy eigenvalues, but also we
restrict the eigenfunctions as well.
One must then also provide the TB fit with a database in which
the energy bands are broken down by their irreducible representation
at each high symmetry point and direction.
In practice this means that the linear combinations of TB basis
functions that have the irreducible representations at each $\textbf{k}$
must be determined.  These linear combinations then ``block-diagonalize''
the Hamiltonian and overlap matrices as described above.

\begin{table}
\caption{\label{tab:bd}
Basis functions used in the block-diagonalization of high-symmetry point
X in the fcc lattice.  Only one of the two partners is shown for the two-
dimensional representations.}
\begin{ruledtabular}
\begin{tabular}{lll}
 $\text{X}_1$    & $A_{1g}$ & $\{ |s\rangle,|d_5\rangle \}$,  \\
 $\text{X}_2$    & $B_{1g}$ & $\{ |d_4\rangle \}$,  \\
 $\text{X}_{2'}$ & $B_{1u}$ & $\{ |f_1\rangle \}$, \\
 $\text{X}_3$    & $B_{2g}$ & $\{ |d_1\rangle \}$,  \\
 $\text{X}_{3'}$ & $B_{2u}$ & $\{ |f_7\rangle \}$, \\
 $\text{X}_{4'}$ & $B_{3u}$ & $\{ |p_3\rangle,|f_4\rangle \}$, \\
 $\text{X}_5$    & $E_g   $ & $\{ (|p_2\rangle+|p_3\rangle)/\sqrt{2} \}$,  \\
 $\text{X}_{5'}$ & $E_u   $ & $\{ |p_1\rangle,|f_2\rangle,|f_5\rangle \}$, 
\end{tabular}
\end{ruledtabular}
\end{table}
The use of symmetry in constraining the fitting process has previously been 
developed by Papaconstantopoulos\cite{papacon86} for $sp^3d^5$ basis.  We have 
found our procedure for using symmetry constraints to be
essential in limiting the parametric phase space and thereby obtaining 
transferable TB parameters.
Here we extend the scheme to include $f$-electrons.
We wish to strongly emphasize how important this procedure has been
in determining an accurate fit, especially with the large parameter set
used when fitting to an $sp^3d^5f^7$ basis.
As an example, in the fcc lattice structure, each energy band at the point X [with 
cartesian coordinates $(0,2\pi/a,0)$] can be decomposed into
different symmetry-adapted combinations of the basis functions,
as shown in Table \ref{tab:bd},
where we have included, for the two-dimensional representations $E_g$
and $E_u$, only one of the two possible partners.  Note that we are
thus able to essentially fit the eigenfunctions as well as the
eigenvalues by this use of the block-diagonalization procedure.  In
the above listing, the first column is simply the conventional band
label, the second the conventional group symbol ($g$ and $u$ are the
usual positive and negative parity labels), and the third column lists
the basis expansion functions belonging to that particular irreducible
representation.

To begin the fitting procedure, it is necessary to make some choice of
initial parameters.  In our experience, the fitting procedure was not
sensitive to the choice of initial parameters, provided that the
block-diagonalization procedure was used as the fit was optimized.

For this reason, we have typically chosen a very simple initial set of
parameters.  We usually set $\lambda=1$, $e_\alpha^0$ near the expected
center of the relevant band and the other $e_\alpha^i=0$ for $i$ = 1,
2, or 3, and the diagonal elements of $s$ =1.  All of the off-diagonal
elements for both the Hamiltonian and overlap matrices are started at
zero.  With this starting set, we typically optimize one of the cubic
structures for 3 or 4 volumes spanning the range of volumes of
interest, and allow the parameters to optimize with block-
diagonalization.  This provides the real starting TB parameters.

The fitting process involves standard nonlinear least squares
algorithms\cite{dennis81}.  Because this type of optimization process
is prone to becoming trapped in local minima, the initial step
described is important in order to provide a good starting point for
the full fitting process.

An alternative starting procedure has been described by
NRL\cite{papacon86} that involves algebraically solving a simplified
nearest-neighbor tight-binding model at high symmetry points for cubic
crystal structures and then comparing the solution to the LAPW bands
at the specified points to determine a resulting reduced set of
parameters.  These are then used as the starting values for the
nearest-neighbor-shell non-hybridizing matrix elements, namely
$\{ss\sigma, pp\sigma, pp\pi, dd\sigma, dd\pi, dd\delta, f\!f\sigma,
f\!f\pi, f\!f\delta, f\!f\phi \}$.  Other matrix elements are started
with zero value.  We have also tried this, more complicated,
procedure.  However, we converged to fits of comparable quality as our 
standard approach (described above).

In the full optimization we include fcc, bcc, and sc crystal
structures in addition to the ground-state crystal structure (if
different) and any other structures of interest.  Because we know the
correct block-diagonalization at high symmetry points and lines for
the cubic phases, keeping these in the fit prevents the minimization
from wandering into unphysical parts of parameter space.  For all
structures and volumes included in the fit, we use both the energy
eigenvalues of the relevant bands as well as the total energy.  During
the minimization procedure, we typically weight the total energies by
a factor of 1,000 to 10,000 times that of individual energy
eigenvalues, and include a penalty factor for singular overlap
matrices.  No block-diagonalization was done for the non-cubic
structures.

\subsection{The Database}

\begin{table}[htb]
\caption{\label{tab:database}
Data used in TB fits, including number of bands fit in each
volume, $N_b$, and the number of volumes fit in each crystal structure,
$N_V$.  Also listed are the average root-mean-square errors in the
fitted energy bands, $\langle\epsilon_b\rangle$, and the average error
in the fitted total energy, $\langle E_T \rangle$.}
\begin{ruledtabular}
\begin{tabular}{llllll}
Element & Structure & $N_V$ & $N_b$ & $\langle \epsilon_b \rangle$ [eV]&
$\langle E_T \rangle$ [eV] \\
Be & fcc & 9 & 3 & 0.73 & 0.0063\\
   & bcc & 7 & 3 & 0.69 & 0.0053\\
   &  sc & 6 & 3 & 1.01 & 0.0034\\
Cu & fcc & 12 & 6 & 0.11 & 0.0018\\
   & bcc &  8 & 6 & 0.12 & 0.0021\\
   &  sc &  9 & 6 & 0.12 & 0.0027\\
Nb & fcc &  7 & 6 & 0.25 & 0.0021\\
   & bcc &  8 & 6 & 0.25 & 0.0018\\
   &  sc &  5 & 6 & 0.29 & 0.0015\\
U  & fcc &  6 & 9 & 0.42 & 0.0105\\
   & bcc &  5 & 9 & 0.36 & 0.0075\\
   &  sc &  6 & 9 & 0.59 & 0.0082\\
   & oC4 & 14 & 18 & 0.39 & 0.0173
\end{tabular}
\end{ruledtabular}
\end{table}
To determine the TB parameters, we have fit to a database of highly
accurate total energies and energy bands from a series of
full-potential linearized augmented plane-wave (FLAPW) calculations
including local orbitals\cite{singh94}.  The data for which each
element was fit consisted of energy bands and total energies for a
series of volumes in various crystal structures.  Table
\ref{tab:database} lists the number of volumes and crystal structure
types for each of the elements studied here, Be, Cu, Nb, and U.

\begin{figure}[htb]
\includegraphics[scale=0.35]{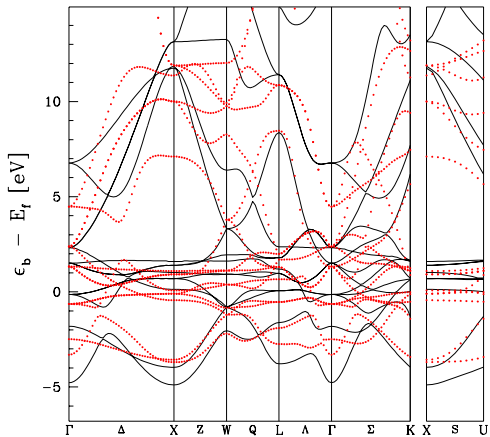}
\label{Ubands}
\caption{Electronic energy bands for fcc U, $a=8.2$, showing the quality
of the TB fit (solid symbols) compared to the FLAPW results.  Note that
the first 9 bands were fitted throughout the IBZ.  Low lying bands are
fit more accurately since they also contribute to the total energy.  Bands
far above the fermi energy were not fit. }
\end{figure}

Also shown in Table \ref{tab:database} is the number of energy bands fit
for $k$-points in the IBZ (47 IBZ $k$-points were used for fcc and bcc
lattices, 56 for sc, and 24 for $\alpha$-U), which were also used in
determining the total energies using a temperature broadening special
point integration (using a broadening of $2$mRy).  The block diagonalization
procedure was carried out at all high-symmetry points as well as the
midpoints of all high symmetry directions in the BZ, which made it 
necessary to break down the FLAPW energy bands by their irreducible
representation at all such points.  A representative plot illustrating
the quality of the TB fit for low lying energy bands is shown in Figure
\ref{Ubands}, in this case for fcc U near the equilibrium volume; note
that we show bands for the fcc structure rather than the ground-state 
structure in order to reduce the complexity of the figure (a similar
figure for $\alpha$-U would have 2 times as many bands).  The fit
quality is quite good for a substantial energy region extending far
above the fermi level.  The highest energy bands involve higher-lying
orbitals that are not included in our tight-binding basis, and hence
the TB fit breaks down at high energy.
For one crystal structure it is often possible to fit the energy bands
so well that the difference between the LAPW and the TB bands are
almost indistinguishable.  However, when all of the different
structures are added into the TB fitting procedure, the overall
quality of the fit to the individual energy bands tends to degrade to
that shown in Figure \ref{Ubands}.  In practice, we have found that
improving the total energy fit lower than 1~mRy requires some
compensating increase in the individual energy band errors.  Ideally
the two sets of errors should collectively decrease (after all, the
total energy is the sum of the occupied band energies), but there may
be some shortcoming in the functional form of the TB parameters that
cause the two sets of errors to compensate for one another rather than
for them to behave collectively.  We hope to explore this issue in
future work by considering alternate functional forms.

These FLAPW results matched very well with a recent {\it ab-initio}
study of the structural properties of U using FLAPW and a
complementary technique using a basis set of Gaussian type
orbitals\cite{boettger99}.  We have neglected spin-orbit coupling for
U, which has little effect on the bulk properties\cite{jones00}.  In a
future paper we will extend this TB technique to include spin orbit
coupling for f-electron materials (application to a three center $sp^3$ basis 
has been explored, for example, in Ref. \cite{lach-hab02}), and apply it to 
heavy elements.

\section{Application to Metallic Elements}

We now apply the TB total energy method to a range metallic elements
from the periodic table.  In order to provide a fair test for the TB
method as an interpolation scheme, we have selected Be to represent
light metals, Cu and Nb for transition metals, and U as a heavy
f-electron metal.  Table \ref{tab:results} list the structural
properties predicted by our TB fits, compared with FLAPW and
experimental results.
\begin{table*}[htb]
\caption{\label{tab:results}
Structural parameters of hcp Be, fcc Cu, bcc Nb, and $\alpha$-U in the
TB model , FLAPW calculation, and experiment (at a temperature where
consistent structural and elastic data was available, noted in
parentheses).  The equilibrium volume is denoted by $v_0$, and the
bulk modulus and its pressure derivative by $B_0$ and $B_0'$.  $y$ is
the atomic position parameter in the oC4 $\alpha$-U structure}
\begin{ruledtabular}
\begin{tabular}{lcccccc}
method & $v_0($\AA$^3)$ & $B_0$(GPa) & $B_0'$ & $c/a$ & $b/a$ & $y$\\
\hline
hcp Be\\
TB     & 8.04 & 127 & 2.0  & 1.565 \\
FLAPW  & 7.90 & 123 & 3.33 & 1.553\\
Expt(100K)\footnote{Ref.~\onlinecite{smith60}}
       & 8.05 & 115\footnote{Ref.~\onlinecite{finkel68}} &      & 1.571\\
\hline
fcc Cu\\
TB     & 11.9 & 139.0 & 5.1 \\
FLAPW  & 11.9 & 140.3 & 5.1 \\
Expt(293K)\footnote{Ref.~\onlinecite{klooster79}}
       & 11.79\footnote{Ref.~\onlinecite{suh88}} & 138 & 5.3  \\
\hline
bcc Nb\\
TB     & 18.1 & 160 & 3.9 \\
FLAPW  & 18.1 & 161 & 3.8 \\
Expt(120K) & 18.02\footnote{Ref.~\onlinecite{roberge75}}
           &171.6\footnote{Ref.~\onlinecite{hayes74}}\\
\hline
$\alpha$-U \\
TB    & $20.24$ & $131$ & $2.8$ & $1.732$ & $2.044$ & $0.1017$ \\ 
FLAPW & $20.14$ & $142$ & $5.0$ & $1.741$ & $2.073$ & $0.0990$ \\
Expt\footnote{Ref.~\onlinecite{barrett63}.}
      & $20.58$ & $136$\footnote{Ref.~\onlinecite{yoo98}} 
			&       & $1.734$ & $2.063$ & $0.1023$ \\
\end{tabular}
\end{ruledtabular}
\end{table*}
In general, the TB fits are very accurate, and highly transferable.
The equilibrium volume, $v_0$, and bulk modulus, $B_0$, as well as the
pressure derivative of the bulk modulus, $B_0'$, were determined by
fitting the various equation of state curves to a second order Birch
equation of state\cite{birch78}.

\textit{ Beryllium.}
The alkaline-earth metal Be belongs to a part of the periodic table
known for nearly free electron behavior, yet the properties of this
metal, including the $c/a$ ratio, are strongly influenced by the
directional character of the interatomic bonding.  In determining our
\begin{figure}[htb]
\includegraphics[scale=0.35]{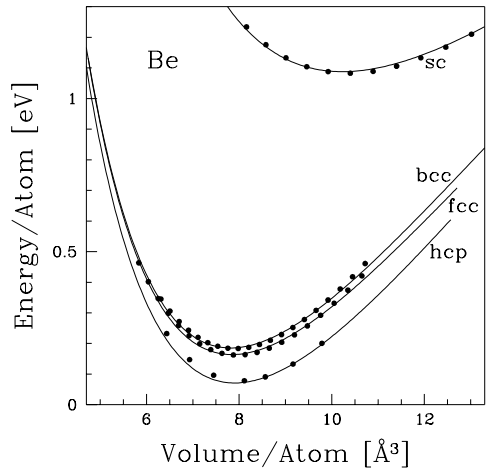}
\caption{\label{fig:Be-eos}
Equation of state curves for various crystal phases of Be,
both in TB (solid symbols), and in FLAPW (solid curves).}
\end{figure}
TB fit for Be, we used a $sp^3d^5$ basis with parameters fit to a
database of 22 different volumes in the fcc, bcc, and sc phases.  It
was necessary to restrict the fit to only the three lowest energy
valence bands ($N_V$ in Table \ref{tab:database}) for Be, as the bands
corresponding to the atomic $3s$ states lie just above the fermi
level, and our basis set has only a single $s$-type basis function (a
fit using an additional $s$ state would be able to fit higher energy
bands).  We found that an accurate model required inclusion of the $d$
orbitals, which are quite close energetically to the occupied s and p
orbitals.  The ground state structural parameters for the Be TB model
are compared with experiment in Table \ref{tab:results}.  Figure
\ref{fig:Be-eos} shows that our TB model successfully predicts the
ground state crystal structure (hcp) in addition to an accurate
prediction of the equilibrium volume and $c/a$ ratio.
The $c/a$ ratio was determined by using a cubic polynomial fit to 
total energy calculations for 8 $c/a$ values around the experimental
value.

\begin{figure}[htb]
\includegraphics[scale=0.35]{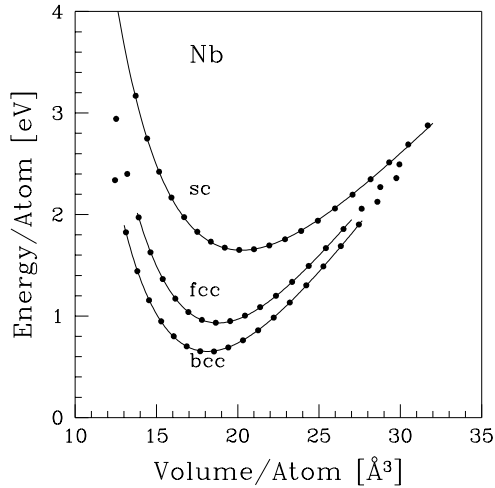}
\caption{\label{fig:Nb-eos}
Equation of state curves for various crystal phases of Nb,
both in TB (solid symbols), and in FLAPW (solid curves).}
\end{figure}

\textit{Copper  and Niobium.}
\begin{figure}[htb]
\includegraphics[scale=0.3]{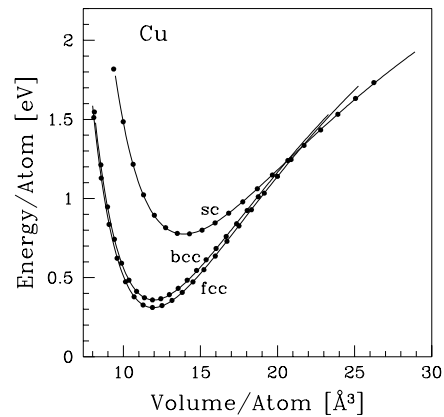}
\caption{\label{fig:Cu-eos}
Equation of state curves for various crystal phases of Cu,
both in TB (solid symbols), and in FLAPW (solid curves).}
\end{figure}
Equation of state plots comparing our TB model for the transition metals 
Cu and Nb are shown in Figures \ref{fig:Cu-eos} and \ref{fig:Nb-eos}.
The predicted structural parameters are given in Table
\ref{tab:results}.  For both transition metals the lowest six energy
bands were fitted throughout the IBZ.  For both elements the ground
state structure is cubic, with a small energy difference between the
face and body-centered structures, which the TB fit reproduces quite
well.  Note the excellent agreement between the TB model and the FLAPW
calculations, as well as the broad range of volumes and energies
fitted.

\textit{Uranium.}
For U, it has been shown that the bulk structural properties are
rather insensitive to the treatment of spin-orbit coupling for the
valence electrons\cite{jones00}, so this element makes a good test
case for the scalar relativistic treatment that we have thus far
implemented in our TB method.  For U, the lowest 9 energy bands per atom 
were fit throughout the IBZ for the cubic crystal structures as well as
the $\alpha$-U structure (oC4).  The atomic position
parameter, $y$, was determined by performing a cubic fit to a six
calculations of the total energy versus $y$.  The ratios $b/a$ and
$c/a$ were then found by using this value of $y$ and performing a set
of total energy calculations at the experimental volume for twenty
values of $b/a$ and $c/a$.  The ratios that minimized the total energy
were then found using a 10 parameter, two dimensional cubic
polynomial.

\begin{figure}[htb]
\includegraphics[scale=0.35]{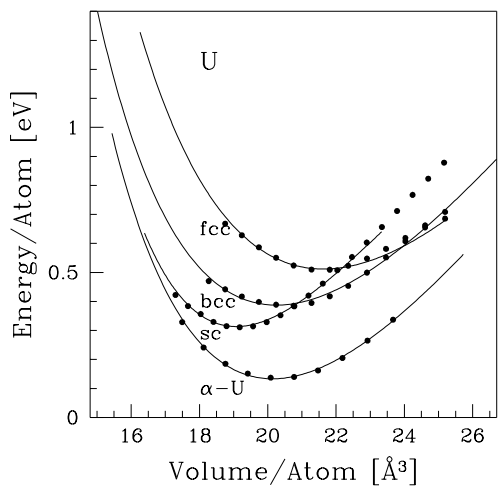}
\caption{\label{fig:eos}
Equation of state curves for various crystal phases of U, both in TB
(solid symbols), and in FLAPW (solid curves).}
\end{figure}

The equation of state for the $\alpha$, fcc, bcc, and sc phases are
shown in Figure \ref{fig:eos}, which shows excellent agreement between
the calculated total energies between FLAPW and TB, thus justifying TB
as an accurate interpolation scheme, even for f-electron materials.
Again the agreement between the TB and FLAPW calculations is excellent.

\section{Conclusions}

We have presented the results of a study of the structural properties
of the metallic elements Be, Cu, Nb, and U using a tight-binding total
energy method.  Fits for the TB method were obtained from highly
accurate FLAPW calculations.  The TB models thus derived are highly
efficient, due to the treatment of valence electrons only, and highly
accurate, proving to be transferable to new structures not included in
the original fit.

We have also shown that the NRL total energy tight-binding parameterization
is highly accurate for f-electron as well as for d-electron systems.
In addition we have demonstrated that forcing the tight-binding fitting
procedure to use the correct wave-function symmetry when fitting to
specific energy eigenvalues at high symmetry $k$ points and lines in
the Brillouin zone leads to highly transferable tight-binding 
parameterizations.  This parameterization seems insensitive to the initial
guess used to start the least-squares minimization procedure.

Transferability is a key issue: do the tight-binding calculations
accurately reproduce what a high-quality first-principles method would
calculate?  Like the answer for any approximation, our conclusions are
mixed.  The results of this paper show that the current NRL
parameterization make it possible to fit the total energy as a
function of volume (for a significantly large range of volumes) for
many different crystal structures.  This is a very significant degree
of transferability.

With respect to other properties, additional detailed results will be
published in future publications for applications to specific
materials.  However, it is possible and relevant to make some
pertinent observations here.  Like any large parameterization, the
properties that can be accurately calculated depend on how well one
has sampled all aspects of the parameterizations in the data base used
to fit the parameters.  Because of the highly complex and nonlinear
relationship between the parameters and the properties, it is always
difficult to guess ahead of time how well one has sampled the full
parameter space.  If the relevant sampling has been done well for the
properties of interest, those properties will be accurately predicted
and compare well with first-principles calculations for the same
properties.

Some preliminary results indicate that elastic constants often are of
about 10-20\% accuracy.  Transferability can be improved by
specifically including in the fit finite distortions of the unit cell
that sample those elastic constants (the distortions should be large
enough to give a significant energy difference on the order of the
difference between other crystal structures life fcc and bcc, yet
small enough so that the energy change is quadratic in energy with the
size of the distortion so that one remains in the elastic regime).
Similarly, preliminary phonon calculations are of a similar degree of
accuracy without any additional fine tuning of the parameters, but can
be significantly improved by adding in a few frozen phonons of finite
amplitude into the fit.

In addition, it is worth pointing out that adding in the diamond
lattice into the fits seems to improve the phonons all by itself
(before adding in frozen phonons, for example).  The diamond lattice
seems to sample other parts of the parameterization that have not been
sampled by the other cubic crystal structures (e.g., with only four
nearest-neighbors, it improves the sample of Eq.\ (\ref{eq:onsite}) to
smaller $\rho$).  As more experience is gained, it may be possible to
find other crystal structures or other properties to include in the
fits that similarly improve transferability.  This is an ongoing work.

Finally, the NRL parameterization is certainly not the only possible
parameterization.  Significant improvement of transferability may be
achieved by finding better functional forms for the hopping, overlap,
and diagonal matrix elements.

For $f$-electron systems, because the rare-earth series of elements
have localized $f$ electrons (beyond Ce), the interesting systems are
the actinide series.  For this series it will be important to include
spin-orbit coupling due to the large charge of the nucleus, especially
for Np and Pu.  We hope to explore addition of spin-orbit coupling to
the tight-binding method for f-electron materials in future work.

To conclude: based on the excellent results of the tight-binding total
energies that we have found for a wide variety of single-element
metals and crystal structures, the tight-binding approach appears
promising for providing highly accurate calculations of complicated
geometries such as defect states and distorted structures, where many
atoms are required in large supercells, and first- principles methods
are prohibitively expensive.

\begin{acknowledgments}
All FLAPW calculations were performed using the Wien97
package\cite{wien97}.  This research is supported by the Department of
Energy under contract W-7405-ENG-36.  Calculations were performed at
the Los Alamos National Laboratory, the Center for Computational
Research at SUNY--Buffalo, and the National Energy Research Scientific
Computing Center (NERSC), which is supported by the Office of Science
of the U.S. Department of Energy under Contract No. DE-AC03-76SF00098.
\end{acknowledgments}

%

\end{document}